\documentclass[11pt,a4paper]{article}

\usepackage{amssymb}
\usepackage{amsmath}
\usepackage[pdftex]{graphicx} 
\usepackage{graphics}
\usepackage{epsfig}
\usepackage{color}
 
 \usepackage{cite}
 
\newcommand{\be}{\begin{equation}}
\newcommand{\ee}{\end{equation}}
\numberwithin{equation}{section}

\newtheorem{theor}{Theorem}
\newtheorem{coro}{Corollary}

\usepackage[hidelinks]{hyperref}
\hypersetup{
  colorlinks   = true, 	
  urlcolor     = red, 	
  linkcolor    = blue, 	
  citecolor   = magenta 	
}

\allowdisplaybreaks

\begin{document}

\title{\bf The one-dimensional Coulomb Hamiltonian: Properties of its Birman-Schwinger operator}

 \author{S. Fassari$^{1}$\footnote{sifassari@gmail.com, 
ORCID: \href{http://orcid.org/0000-0003-3475-7696}{0000-0003-3475-7696}}, 
 M. Gadella$^{2,3}$\footnote{manuelgadella1@gmail.com, 
 ORCID: \href{http://orcid.org/0000-0001-8860-990X}{0000-0001-8860-990X}}, 
 J.T. Lunardi$^4$\footnote{jttlunardi@uepg.br, 
 ORCID: \href{https://orcid.org/0000-0001-7058-9592}{0000-0001-7058-9592}}, 
L.M. Nieto$^{2,5}$\footnote{luismiguel.nieto.calzada@uva.es, 
ORCID: \href{http://orcid.org/0000-0002-2849-2647}{0000-0002-2849-2647}}, 
and 
F. Rinaldi$^{1}$\footnote{f.rinaldi@unimarconi.it, 
ORCID: \href{http://orcid.org/0000-0002-0087-3042}{0000-0002-0087-3042}}
\\ \\
\small  
$^1$ Dipartimento di Scienze Ingegneristiche, \\
\small  
Univ. degli Studi Guglielmo Marconi, Via Plinio 44, I-00193, Rome, Italy
\\
\small  
  $^2$ Departamento de F\'{\i}sica Te\'orica, At\'omica y Optica,  
  U. de Valladolid, 47011 Valladolid, Spain
\\ 
\small  
$^3$   IMUVA, Universidad de Valladolid, 47011 Valladolid, Spain
\\ 
\small  
 $^4$ Departamento de Matemática e Estatística,
Universidade Estadual de Ponta Grossa \\
\small  
Av. Carlos Cavalcanti, 4748, 84030-900, Ponta Grossa, Brazil
\\
\small  
$^5$ Laboratory for Disruptive Interdisciplinary Science (LaDIS),  
 U. de Valladolid, 47011 Valladolid, Spain
  }

\maketitle

\begin{abstract}
\noindent
The objective of the present paper is to study in detail the properties of the Birman-Schwinger operator for a self-adjoint realisation of the one-dimensional Hamiltonian with the Coulomb potential, both when the Hamiltonian is defined only on $\mathbb R_+$ and when it is defined on the whole real line. In both cases, we rigorously show that the Birman-Schwinger operator is Hilbert-Schmidt, even though it is not trace class, a result that, to the best of our knowledge, has not yet been achieved in the existing mathematically oriented literature on the one-dimensional hydrogen atom. Furthermore, in both cases we have considered some Hamiltonians depending on a positive parameter approximating the Hamiltonian with the Coulomb potential and proved the convergence of the Birman-Schwinger operators of such approximations to the corresponding one of the Hamiltonian with the Coulomb potential as the parameter goes to zero. The latter result implies the norm resolvent convergence of the approximating Hamiltonians to that of the Hamiltonian with the Coulomb potential.

\end{abstract}

\vspace{2pt}

\vspace{2pt}

\noindent
AUTHOR CONTRIBUTIONS: 
The authors state that the research was carried out in conjunction with one another and with equal accountability. The final text was reviewed and approved by all of the authors.
\vspace{2pt}

\noindent
CONFLICT OF INTEREST STATEMENT: 
The writers state unequivocally that they do not have any conflicts of interest.
\vspace{2pt}

\noindent
PAPER SUBMITTED TO: 
{\it Mathematical Methods in the Applied Sciences}

\newpage

{\color{blue}\it We wish to dedicate this article to the memory of Prof. Dr. Martin Klaus whose remarkable body of work on both the spectral and the scattering analysis of Schr\"odinger operators has been a constant source of inspiration for us.}

\section{Introduction}

The present paper is intended to be a new contribution of Operator Theory to mathematical problems and models that take their origin in Quantum Mechanics. In this context, we use self-adjoint operators to represent quantum observables. In particular, we consider that the Hamiltonian, which defines the model under study, is always a self-adjoint operator on a Hilbert space of the type $L^2(\Delta)$. One dimensional models are those in which $\Delta$ is either the real line $\mathbb R$ or a closed subspace thereof. One-dimensional models are used in Quantum Physics to study properties of the systems due to their greater simplicity than others in higher dimensions and because they are often solvable. 

In the study of Quantum Mechanics,  Hamiltonians are usually chosen self-adjoint\footnote{Here we do not wish to discuss the case of Hamiltonians which are not self-adjoint but PT-invariant.} for various reasons that include: i.) the possibility of defining unitary time evolution operators, so that probabilities are conserved with time; ii.) conservation of probability current, an important physical property, or; iii.) reality of the spectrum (continuous spectra of symmetric operators which are not self-adjoint contain points belonging to $\mathbb C {\setminus }\mathbb R$). We consider Hamiltonians which can be written as a sum of two terms, the former (resp. the latter) being the kinetic (resp. potential) energy plus a potential, i.e., $H=-\nabla ^2-V, V\geq 0$, where we have restricted ourselves to the case of nonpositive potentials due to the attractive nature of the Coulomb potential in the hydrogen atom. 

The current literature on Mathematical Physics as well as on Quantum Mechanics discusses  a great variety of Hamiltonian models for which the potential, or a component of the potential, is supported on a manifold having a smaller dimension than that of the ambient space. These potentials mimic forces with great strength placed on a very thin region of the space. Other models include potentials with other kinds of singularities. In general, such potentials are called point potentials, localised potentials or even singular potentials. Usually, their Hamiltonians are the sum of the $n$-th ($n\ge 1$) dimensional Laplacian $-\nabla^2$  plus a point or singular interaction, typically a Dirac delta or a sum of Dirac deltas, and possibly other terms. 

The question on how to determine self-adjoint versions of such Hamiltonians is often not simple. One usual procedure is the following: One starts with a {\it non-perturbed} Hamiltonian, $H_0$, which is a symmetric operator with equal deficiency indices on a given domain, and then find its self-adjoint extensions and their domains. These new self-adjoint determinations of the original symmetric operator may have formal differences with respect to the former, which may be physically relevant (for instance having a different spectrum), and which may manifest themselves by the addition of one or more of these ``singular'' potential terms. 

An example extensively studied in the literature is the one-dimensional Laplacian $-d^2/dx^2$ defined, say, on the Sobolev space $W_2^2(\mathbb R {\setminus }\{0\})$ as a subspace of $L^2(\mathbb R)$. This Laplacian is symmetric with deficiency indices $(2,2)$, so that it admits a four parametric family of self-adjoint connected extensions plus a two parametric family of separated (disconnected) estensions. In this case,  all self-adjoint extensions and, hence, the families of all perturbing potentials, are fixed through matching conditions at the origin. For instance, one of these extensions can be characterised by the upsurge of a potential of the type $V(x)= \alpha\, \delta(x)$, where $\alpha \ne 0$ is a real number and $\delta(x)$ is the renowned Dirac distribution. This procedure may be extended to a countably infinite set of points along the real line, without limit points \cite{K}. 

Although this procedure of finding self-adjoint extensions through matching conditions may seem rather simple in principle, it is not so in the general case. Some extensions cannot be obtained in this way and require some more sophisticated  and complicated constructions via regularisation processes. This is, for instance, the typical situation when one departs from a Hamiltonian $H_0=-\nabla^2$ with dimension $d>1$ (on an appropriate domain). However, it may also be the case where $H_0$ is a one-dimensional pseudo-differential operator \cite{AK,EGU}. These ideas have often been the source of new mathematics \cite{LL}.

As mentioned before, another type of non-regular potentials are those exhibiting singularities. This is the case of the example under our study: the so-called 
one-dimensional Coulomb potential. Although the self-adjoint extensions of this Hamiltonian have been studied elsewhere \cite{FLM95,CES,CAL}, it is certainly worth studying them as they might represent a laboratory of new mathematical techniques meant to find properties of self-adjoint determinations of symmetric operators, the spectral ones being the most relevant ones. In the present paper, we use techniques based on the use of the Birman-Schwinger  B-S operator \cite{Klaus82}, in order to obtain spectral properties of the one-dimensional Coulomb Hamiltonian and, at the same time, we adopt a rigorous approach to some of the standard regularisations used for the Coulomb potential. 

Then, our point of departure is given by the following formal Hamiltonian:
\begin{equation}\label{1}
H:=  H_0-V= -\frac{d^2}{dx^2} - \frac {\lambda}{|x|}\,, \qquad x\ne 0\,, \quad \lambda>0\,.
\end{equation} 
Contrary to the $d$-dimensional case ($d=2,3$), where the singularity $-\lambda/r$ is balanced by  the Jacobian resulting from the change from Cartesian to spherical coordinates, in one dimension there is no mechanism to compensate the singularity due to the Coulomb potential{\footnote{Since the potential in (\ref{1}) is not a solution for the one-dimensional Poisson equation with a point source, it is not the  proper Coulomb potential in one dimension. However, we follow the usage in the literature and refer to it as the one-dimensional Coulomb potential. The term  {\it 1-D hydrogen atom} is also used in the literature in order to refer to the system with the Hamiltonian (\ref{1}).}.  Therefore, the potential term in \eqref{1} is far more singular than its counterpart in higher dimensions and, as a consequence, the above Hamiltonian has a richer family of self-adjoint extensions.

The issue regarding the self-adjoint realisation of the Hamiltonian \eqref{1} took centre stage in a remarkable number of articles such as \cite{FLM95,KUR96,FLM97,KUR97,M,DDL,G1,G2}. In particular, \cite{FLM95} includes both the functional analytic and Feynman's path integral approach in relation to the quest for appropriate realisations of \eqref{1}. In \cite{CES}, we find a detailed discussion on the self-adjointness of these realisations. Nevertheless, the mathematical analysis of the determinations of \eqref{1} is not yet over, as attested by two recent publications \cite{SHU,ZHA}. 

In the present context we should also mention the crucial contribution by Klaus \cite{KLA} who, by improving Gesztesy's earlier result \cite{Fritz}, proved rigorously that, under suitable assumptions on the smearing potential $-\lambda V_{\varepsilon}$ approximating the one-dimensional Coulomb potential  $-\lambda/|x|$, the Hamiltonian $ -\frac{d^2}{dx^2} -\lambda V_{\varepsilon}$ converges in the norm resolvent sense (see \cite{RSI}) to the Hamiltonian $ -\left[\frac{d^2}{dx^2} \right]_D -\lambda/|x|$, which is bounded from below, $ -\left[\frac{d^2}{dx^2} \right]_D$ being the one-dimensional Laplacian with the Dirichlet boundary condition at the origin. Given that the compactness of the    B-S   operator pertaining to the Hamiltonians involved plays a crucial role throughout \cite{KLA}, it is actually the latter paper that motivated us to investigate in detail the B-S operator pertaining to  $ -\left[\frac{d^2}{dx^2} \right]_D -\lambda/|x|$ in the present work, which led us to establish that such a B-S operator belongs to the Hilbert-Schmidt class. We were further motivated to revisit the one-dimensional hydrogen atom by a spectral implication of Klaus' aforementioned norm resolvent convergence: as aptly pointed out by Klaus, due to the principle of noncontraction of the spectrum, {\it which holds exclusively under norm resolvent convergence} (see \cite{RSI}), while the eigenvalue pertaining to the symmetric ground state eigenfunction of the approximating Hamiltonians disappears in the limit $\varepsilon \rightarrow 0_+$, the one of the first excited state, having an antisymmetric eigenfunction, converges to that pertaining to the antisymmetric eigenfunction of the ground state of $ -\left[\frac{d^2}{dx^2} \right]_D -\lambda/|x|$. This is a manifestation of the change of symmetry of the ground state eigenfunction at a critical value of the parameter labelling a given Hamiltonian, a phenomenon occurring in the coalescence problem involving a pair of identical attractive nonlocal point perturbations $|\delta'><\delta'|$ symmetrically situated about the origin, recently investigated in \cite{EPJP24}, where the crucial parameter is the distance between the centres of the point interactions. It may be worth recalling that such a symmetry reversal of the ground state eigenfunction manifests itself in at least two other situations involving an attractive nonlocal point perturbation $|\delta'><\delta'|$, namely: 

i.)  the Hamiltonian of the one-dimensional harmonic oscillator perturbed by the attractive nonlocal point perturbation $-\beta |\delta'><\delta'|$ centred at the origin, $\beta>0$ being the coupling parameter resulting from the renormalisation procedure, exhibiting the aforementioned change of symmetry when $\beta$ exceeds the critical value $\beta_0=\frac {\Gamma(1/4)}{2\Gamma(3/4)}\approx 1.47934$, as shown in \cite{AFR4,AFR5,FGGNR}; 

ii.) the Hamiltonian of the one-dimensional conic or V-shaped oscillator perturbed by the attractive nonlocal point perturbation $-\beta |\delta'><\delta'|$ centred at the origin, $\beta>0$ being the coupling parameter resulting from the renormalisation procedure, exhibiting the aforementioned change of symmetry when $\beta$ exceeds the critical value $\beta_0=-\frac {Ai(0)}{Ai'(0)}\approx  1.37172$, as shown in \cite{FGGN,FGGNR}. 

Some regularisation attempts of \eqref{1} have been undertaken \cite{G1,G2}. Typically, one replaces the term $-\lambda/|x|$ by means of a regular potential symmetric with respect to the origin with a minimum at $x=0$ and then takes the limit as the minimum goes to minus infinity \cite{LOU,LOU1,OV}. Another alternative consists in replacing this term $-\lambda/|x|$ by either $-\lambda/(|x|+\varepsilon)$ or 
$\lambda/\sqrt{x^2+\varepsilon^2}$, $\varepsilon>0$, or a similar one. Then,  take the  limit as $\varepsilon \to 0^+$. In some other cases, it is customary to add a cut-off to the Coulomb term \cite{LOU1,RUE}.

In this paper, we shall focus our attention on a particular self-adjoint determination of \eqref{1} of particular interest in Physics and analyse relevant properties concerning its related B-S operator. 

The   B-S principle has been discussed in  \cite{Klaus82,RSIV}. These works make use of the  B-S  operator which has its own interest from the mathematical point of view and, in addition, gives rise to an approximate determination of the ground state and the lowest lying excited states for non exactly solvable one-dimensional Hamiltonians which cannot be easily treated by other means. The main objective of the present contribution is the study of the    B-S  operator for the present model. It comes directly from the Schr\"odinger equation for an operator $H=H_0-V$ like the one in \eqref{1}, where $H_0$ is usually (although not necessarily) the kinetic term and $V$ is the potential. This Schr\"odinger equation can be written in the form
\begin{equation}\label{2}
(H_0-E)\psi=V\psi\,, \qquad E<0 \,.
\end{equation}
Given that $V \geq 0$, so that its square root is well defined, then, after \eqref{2}
\begin{equation}\label{3}
(H_0-E) V^{-1/2} V^{1/2}\psi = V^{1/2} V^{1/2} \psi\,.
\end{equation}
By setting $\chi:= V^{1/2}\psi$, then \eqref{3} becomes
\begin{equation}\label{4}
\chi = V^{1/2} (H_0-E)^{-1} V^{1/2} \chi = B_{BS}(E) \chi\,.
\end{equation}
The operator
\begin{equation}\label{5}
B_{BS}(E) := V^{1/2} (H_0-E)^{-1} V^{1/2}
\end{equation}
is the renowned    B-S  operator. Before moving forward, we wish to remark that, as attested by a recent contribution to the topic \cite{HRS}, the B-S operator has also become increasingly relevant in the spectral analysis of a Hamiltonian not related to the Schr\"odinger equation, namely the one arising in relation to the linear Bardeen-Cooper-Schrieffer (BCS) equation, the crucial equation in the theory of superconductivity. 

Going back to our context as described so far, we note that, whenever $V$ is a multiplication operator $V(x)$, $B_{BS}(E)$ admits, in general, an integral kernel, so that 
\begin{equation}\label{6}
[B_{BS}(E) \psi](y)= \int K_{BS}(x,y;E)\,\psi(x) dx\,.
\end{equation}

Discrete solutions of the Schr\"odinger equation \eqref{2} on $E$ give the point spectrum of the Hamiltonian $H=H_0-V$. From \eqref{4}, one immediately sees that these energy levels are given by those values of $E$  for which the operator $B_{BS}(E)$ has an eigenvalue equal to one. In general, this does not provide an accurate method in order to solve \eqref{2}, even though the lowest eigenvalues can be determined within a reasonable approximation. When $B_{BS}(E)$ is a positive integral operator of the form \eqref{6}, it is well known (see \cite{RSIII}) that it is trace class if and only if
\begin{equation}\label{7}
\int K_{BS}(x,x;E)\, dx\ <\infty\,.
\end{equation} 
In this case, as shown in \cite{RSIV}, the energy levels are the solutions of an equation in terms of a Fredholm determinant involving the B-S operator, namely $\det [I-B_{BS}(E)]=0$. In general, this determinant is written as a series, so that only approximate solutions can be found, in general in terms of a given parameter \cite{RSIV,SimonAP76,AGHH,SF89,SF90}. If instead $B_{BS}$ were Hilbert-Schmidt, then a similar result could be obtained by replacing the Fredholm determinant by means of the so-called modified Fredholm determinant  {\cite{SimonAP76,AGHH,Sym21,Math22}.

This is exactly what happens in relation to the determinant associated to the operator $B_{BS}$ pertaining to $H$ in \eqref{1}, as we shall show in the present paper. With the aid of this fact we shall revisit the situation that emerges when we replace the Coulomb term in \eqref{1} by   $-\lambda/(|x|+\varepsilon)$ or $-\lambda/\sqrt{x^2+\varepsilon^2}$ or even other similar potentials with given properties and, at a later stage, take the limit $\varepsilon \to 0^+$. By imposing the Dirichlet boundary conditions at the origin on the one-dimensional Laplacian $-d^2/dx^2$, we see that all the    B-S   operators corresponding to all these Hamiltonians converge, in the Hilbert-Schmidt norm, to the   B-S   operator for \eqref{1}. It is interesting that this convergence implies, in the present case, the convergence of the Hamiltonians in the norm resolvent sense. 

The paper is organised as follows: in Section 2, we have obtained the   B-S   operator for the given self-adjoint determination of \eqref{1}. We have studied two cases, one in which \eqref{1} is defined on the entire real line and the other only on the half-line $\mathbb R^+\equiv [0,\infty)$. In both cases, we have imposed the Dirichlet boundary conditions at the origin. We have proved that the corresponding   B-S   operators  in both cases are Hilbert-Schmidt but not trace class. In Section 3, we have considered a class of approximations to the Hamiltonian \eqref{1} and shown the convergence in the Hilbert-Schmidt norm of the   B-S   operators for these approximations to the B-S operator of the chosen self-adjoint determination of \eqref{1}.  Furthermore, by comparing our approximating potentials and those used by Klaus in \cite{KLA}, we have pointed out why his result holds under our upper bound for the approximating potentials.
Finally, we have added some concluding remarks and 
an Appendix with some comments referring to some   B-S    operators similar to the ones for \eqref{1}.

\section{The Birman-Schwinger operator}

Let us consider the Laplacian operator $-d^2/dx^2$ on $L^2(\mathbb R^+)$, where $\mathbb R^+ \equiv [0,\infty)$. A domain of self-adjointness of this operator is the Sobolev space $W_2^2$ with $f(0)=0$, the latter condition being the Dirichlet boundary condition at the origin. Henceforth, we denote this self-adjoint determination as $\left( -\frac{d^2}{dx^2} \right)_{D^+}$. Its Green function (the kernel of the resolvent operator) is well known (see, for example, \cite{De,EPJP21}) and is given for $E<0$ by
\begin{equation}\label{2.1}
\left[ \left( -\frac{d^2}{dx^2} \right)_{D^+} +|E| \right]^{-1}\!\!\!(x,y) = \frac{e^{-|E|^{1/2}|x-y|}-e^{-|E|^{1/2}(x+y)}}{2|E|^{1/2}}\,.
\end{equation}

As pointed out in \cite{KLA}, see also \cite{Klaus82,De,Sim77}, its counterpart for the self-adjoint realisation of $\left( -\frac{d^2}{dx^2} \right)_{D}$ is given by:
\begin{equation}\label{2.2}
\left[ \left( -\frac{d^2}{dx^2} \right)_{D} +|E| \right]^{-1}\!\!\!(x,y) = \frac{e^{-|E|^{1/2}|x-y|}-e^{-|E|^{1/2}(|x|+|y|)}}{2|E|^{1/2}}\,.
\end{equation}

A result that we do not prove here, since it comes easily from some well-known results established \footnote{In particular Theorem 3.1.2 in Section I.3 and Theorem 3.1.1.} in \cite{AGHH}, the self-adjoint realisation of $\left( -\frac{d^2}{dx^2} \right)_{D}$, with the Dirichlet condition $f(0)=0$, admits the following split:
\begin{equation}\label{2.3}
\left( -\frac{d^2}{dx^2} \right)_{D}=\left( -\frac{d^2}{dx^2} \right)_{D^{-}}\, \bigoplus \, \left( -\frac{d^2}{dx^2} \right)_{D^{+}} \,,
\end{equation}
where we define $\left( -\frac{d^2}{dx^2} \right)_{D^{-}}$ exactly as $\left( -\frac{d^2}{dx^2} \right)_{D^{+}}$ just by replacing $L^2(\mathbb R^+)$ by $L^2(\mathbb R^-)$, where $\mathbb R^+ \equiv [0,\infty)$ and $\mathbb R^-\equiv (-\infty,0]$. Thus, we may say that the Dirichlet condition at the origin decouples $\left(-\infty,0\,\right)$ and $\left(0,+\infty \right)$.

Next, we consider the restriction to $L^2(\mathbb R^+)$ of \eqref{1} with the same boundary conditions for the Laplacian. This gives the following operator:
\begin{equation}\label{2.4}
\left( -\frac{d^2}{dx^2} \right)_{D^+} - \frac{\lambda}{x}\,, \qquad x>0\,, 
\end{equation} 
for which the kernel of its   B-S    operator, $B^+_{BS}(E)$, is given by
\begin{eqnarray}\label{2.5}
K^+_{BS}(x,y,|E|) \!\!&\!=\!&\!\! \lambda \frac{1}{x^{1/2}}\, \left[ \left( -\frac{d^2}{dx^2} \right)_{D^+} +|E| \right]^{-1}\!\!\! (x,y) \ \frac{1}{y^{1/2}} \nonumber \\ [2ex]  
\!\!&\!=\!&\!\! \lambda \frac{1}{x^{1/2}}\,  \frac{e^{-|E|^{1/2}|x-y|}-e^{-|E|^{1/2}(x+y)}}{2|E|^{1/2}} \, \frac{1}{y^{1/2}}\,.
\end{eqnarray}

We wish to underline the difference between the Hamiltonian in \eqref{2.4} and the one in \eqref{1}. The former, \eqref{2.4}, is defined on the positive semiaxis, $x>0$, in $x$-space, while \eqref{1} is valid for $x\in \mathbb R$. 

As proved in \cite{CAL,CES}, the operator \eqref{1} with the Dirichlet boundary conditions at the origin is well defined. The kernel of its   B-S  operator takes the following form:
\begin{eqnarray}\label{2.6}
K_{BS}(x,y;|E|) \!\!&\!=\!&\!\! \lambda\ \frac{1}{|x|^{1/2}}\,  \left[ \left( -\frac{d^2}{dx^2} \right)_{D} +|E| \right]^{-1}\!\!\! (x,y) \ \frac{1}{|y|^{1/2}} \nonumber \\ [2ex] 
 \!\!&\!=\!&\!\!  \lambda\ \frac{1}{|x|^{1/2}}\,  \frac{e^{-|E|^{1/2}|x-y|}-e^{-|E|^{1/2}(|x|+|y|)}}{2|E|^{1/2}} \, \frac{1}{|y|^{1/2}}\,, 
\end{eqnarray}
where we have omitted the superindex $+$, whenever it appears in \eqref{2.6}, in order to properly distinguish between both kernels.

Neither   B-S  operator represented by the kernels \eqref{2.5} and \eqref{2.6} is trace class. For instance, as a direct consequence of \eqref{2.5}, we have that
\begin{equation}\label{2.7}
\int_0^\infty K^+_{BS}(x,x;|E|)\, dx = \frac{\lambda}{2|E|^{1/2}} \int_0^\infty \frac{1-e^{-2|E|^{1/2}x}}{x}\,dx = \infty\,,
\end{equation}
the integral being clearly divergent since the integrand behaves at infinity like $1/x$. Note that the function under the integral sign has a finite limit at the origin. The same obviously holds for \eqref{2.6}. Observe that, in both cases, the divergence of the integral is logarithmic, so that one may well conjecture that both   B-S  operators are Hilbert-Schmidt. In order to prove this statement, let us write the kernel \eqref{2.5} as follows:
\begin{eqnarray}\label{2.8}
K^+_{BS}(x,y;|E|) \!\!&\!=\!&\!\! \lambda\, \frac{e^{-|E|^{1/2}x}}{x^{1/2}}\, \frac{e^{|E|^{1/2}[x+y-|x-y|]}-1}{2|E|^{1/2}}\, \frac{e^{-|E|^{1/2}y}}{y^{1/2}} \nonumber \\ [2ex] 
\!\!&\!=\!&\!\!\lambda\, \frac{e^{-|E|^{1/2}x}}{x^{1/2}}\, \frac{e^{ 2|E|^{1/2}\,\min(x,y)}-1}{2|E|^{1/2}}\, \frac{e^{-|E|^{1/2}y}}{y^{1/2}}\,.
\end{eqnarray}
Analogously, we may write the kernel \eqref{2.6} as follows:
\begin{eqnarray}\label{2.9}
K_{BS}(x,y;|E|) \!\!&\!=\!&\!\!\lambda\, \frac{e^{-|E|^{1/2}|x|}}{|x|^{1/2}}\, \frac{e^{|E|^{1/2}(|x|+|y|-|x-y|)} -1}{2|E|^{1/2}}\,  \frac{e^{-|E|^{1/2}|y|}}{|y|^{1/2}} \nonumber \\  [2ex] 
\!\!&\!=\!&\!\! \lambda\, \frac{e^{-|E|^{1/2}|x|}}{|x|^{1/2}}\, \frac{e^{2|E|^{1/2}\min(|x|,|y|)} -1}{2|E|^{1/2}}\,  \frac{e^{-|E|^{1/2}|y|}}{|y|^{1/2}} \,.
\end{eqnarray}

Our next goal is to show that \eqref{2.5} is the kernel of a positive Hilbert-Schmidt operator. To this end, it is sufficient to prove that the following integral converges:
\begin{eqnarray}\label{2.10}
\int_0^\infty \int_0^\infty [K^+_{BS}(x,y,|E|)]^2  dx dy \!\!&\!=\!&\!\! 
\frac{\lambda^2}{4|E|} \int_{0}^{\infty}  \int_{0}^{\infty} e^{- 2|E|^{1/2}x} \left[\frac {e^{2|E|^{1/2} \min(x,y)}-1}{x^{1/2}y^{1/2}}\right]^2 
e^{- 2|E|^{1/2}y}\, dx  dy \nonumber \\ [2ex] 
\!\!&\!=\!&\!\!  \frac{\lambda^2}{4|E|}\, \int_{0}^{\infty}  \int_{0}^{\infty} e^{- x}\, \left[\frac {e^{\min(x,y)}-1}{x^{1/2}y^{1/2}}\right]^2\, 
e^{- y}\, dxdy = { \frac{\lambda^2}{4|E|}A},
\end{eqnarray}
with $A$ in \eqref{2.10} defined as follows:
\begin{equation}\label{2.11}
 A= 2 \int_{0}^{\infty} I(x) \frac {e^{-x}}{x}\, dx\,,
\end{equation}
where
\begin{eqnarray}\label{2.12}
I(x) =  \int_{0}^{x} \frac {e^{-y}}{y}\, \left(e^{y}-1\right)^2 \, dy 
=4 \int_{0}^{x}  \frac {\sinh^2(y/2)}{y} \, dy=\, 4 \int_{0}^{x/2}  \frac {\sinh^2 y}{y} \, dy\,.
\end{eqnarray}
By using the expansion
\begin{equation}\label{2.13}
\sinh^2y=\sum_{\ell=1}^{\infty}\frac {2^{2\ell-1}}{(2\ell)!}\, y^{2\ell}\,,
\end{equation}
in the last expression in \eqref{2.12} we obtain:
\begin{eqnarray}\label{2.14}
I(x) \!\!&\!=\!&\!\! 4 \int_{0}^{x/2}  \frac 1{y} \, \left(\sum_{\ell=1}^{\infty}\frac {2^{2\ell-1}}{(2\ell)!}\, y^{2\ell}\right) dy= 4 \sum_{\ell=1}^{\infty}\frac {2^{2\ell-1}}{(2\ell)!}\, \int_{0}^{x/2} y^{2\ell-1} \, dy  \nonumber \\ [2ex]  
\!\!&\!=\!&\!\!
4 \sum_{\ell=1}^{\infty} \frac {2^{2\ell-1}}{(2\ell)!}\, \left[\frac {y^{2\ell}}{2\ell} \right]_0^{\frac {x}2} =
 2 \sum_{\ell=1}^{\infty} \frac {2^{2\ell}}{2\ell(2\ell)!}\, \left(\frac {x}2\right)^{2\ell}= \sum_{\ell=1}^{\infty} \frac {x^{2\ell}}{\ell(2\ell)!} \,.
\end{eqnarray}
Hence, we can replace $I(x)$ in \eqref{2.11} by means of the last series in \eqref{2.14}, so that we get
\begin{eqnarray}\label{2.15}
A \!\!&\!=\!&\!\!   2   \int_{0}^{\infty} I(x)\, \frac {e^{-x}}{x}\, dx=
  2 \int_{0}^{\infty} \left(\sum_{\ell=1}^{\infty} \frac {x^{2\ell}}{\ell(2\ell)!}\right)\, \frac {e^{-x}}{x}\, dx = 2  \sum_{\ell=1}^{\infty} \frac 1{\ell(2\ell)!} \int_{0}^{\infty} x^{2\ell-1} e^{-x}\, dx \nonumber \\ [2ex] 
\!\!&\!=\!&\!\!  2  \sum_{\ell=1}^{\infty} \frac 1{\ell(2\ell)!}\, \Gamma(2\ell) =  \sum_{\ell=1}^{\infty} \frac 1{\ell^2(2\ell)!}\, 2\ell\,\Gamma(2\ell) = \sum_{\ell=1}^{\infty} \frac 1{\ell^2} = \zeta(2)= \frac {\pi^2}6 \,,
\end{eqnarray}
where $\zeta(z)$ is the Riemann zeta function \cite {Arf}. Therefore,

\begin{equation}\label{2.16}
\int_0^{\infty} \int_0^\infty [K^+_{BS}(x,y,|E|)]^2  dx dy =\frac{\lambda^2\, \pi^2}{24|E|}
\end{equation}

After inspecting \eqref{2.9}, we note that the integral kernel $K_{BS}(x,y;|E|)$ is symmetric with respect to the origin in both variables $x$ and $y$, so that:
\begin{equation}\label{2.17}
\int\int_{\mathbb R^2} [K_{BS}(x,y;|E|)]^2\,dx\,dy = 4\int_0^\infty \int_0^\infty [K^+_{BS}(x,y;|E|)]^2\, dx\, dy = \frac{\lambda^2\, \zeta(2)}{|E|} = \frac{\lambda^2\, \pi^2}{6|E|}\,.
\end{equation}
This confirms that both    B-S   operators, $B_{BS}(E)$ and $B^+_{BS}(E)$ are Hilbert-Schmidt. In addition,  we have calculated their Hilbert-Schmidt norms, $||-||_{T^2}$, given by, respectively,
\begin{equation}\label{2.18}
||B_{BS}||_{T^2}= \sqrt{\frac{1}{6|E|}} \, \lambda \, \pi\,, \qquad ||B^+_{BS}||_{T^2} = \sqrt{\frac{1}{6|E|}} \,\lambda\,  \frac \pi 2\,.
\end{equation}
Now, let us go back to \eqref{2} and write it in the following manner \cite{FNR}:
\begin{equation}\label{2.19}
(H_0-E)^{1/2} \psi= (H_0-E)^{-1/2} V (H_0-E)^{-1/2} (H_0-E)^{1/2}\psi\,.
\end{equation}
Then if $\chi:= (H_0-E)^{1/2} \psi$, equation \eqref{2.19} becomes
\begin{equation}\label{2.20}
\chi =  (H_0-E)^{-1/2} V (H_0-E)^{-1/2} \, \chi\,.
\end{equation}
After \eqref{2.20} and in full analogy with what happens in the case of the   B-S   operator, the energy values for the bound states of \eqref{2} are those values of $E$ for which the operator 
\begin{equation}\label{2.21}
R(E) :=  (H_0-E)^{-1/2} V (H_0-E)^{-1/2}
\end{equation} 
has an eigenvalue equal to one. As is well known \cite{Klaus82,Deift,Klaus79}, the operators $B_{BS}(E)$ defined in \eqref{5} and $R(E)$ are isospectral. The latter is also a positive integral operator which in our case has the following kernel:
\begin{equation}\label{2.22}
R(x,y;|E|) = \lambda \int_{-\infty}^\infty  \left[ \left( -\frac{d^2}{dx^2} \right)_D + |E|  \right]^{-\frac12} \!\!\!(x,z) \ \frac{1}{|z|} \, \left[ \left( -\frac{d^2}{dx^2} \right)_D + |E|  \right]^{-\frac12}\!\!\! (z,y)\ dz\,,
\end{equation}
for which its restriction to $L^2(\mathbb R^+)$ has its integral kernel given by

\begin{equation}\label{2.23}
R^+(x,y;|E|) = \lambda \int_{0}^\infty  \left[ \left( -\frac{d^2}{dx^2} \right)_{D^+} + |E|  \right]^{-\frac12} \!\!\! (x,z) \ \frac{1}{z} \, \left[ \left( -\frac{d^2}{dx^2} \right)_{D^+} + |E|  \right]^{-\frac12} \!\!\!(z,y)\ dz \,,
\end{equation}
as becomes obvious after \eqref{2.22}. If we call $R^+(E)$ the operator on $L^2(\mathbb R^+)$ with the latter integral kernel, we conclude that the Hilbert-Schmidt norms of both, $R(E)$ and $R^+(E)$ coincide with those of $B_{BS}(E)$ and $B^+_{BS}(E)$ as in \eqref{2.18}, respectively, due to the isospectrality of $B_{BS}(E)$ and $R(E)$ on one side and of $B^+_{BS}(E)$ and $R^+(E)$ on the other. It is worth pointing out that, due to the aforementioned isospectrality, operators like $R(E)$ and $R^+(E)$ can be used in place of their respective B-S operators inside the Fredholm determinant whenever the trace class condition is met, that is to say the eigenenergies are also given by the solutions of $\det [I-R(E)]=0$ \cite{FNR,F96,FI97,Acta}.
As a consequence, we have the following result:
\begin{theor}
 For any $\lambda>0,\, E< 0$, the isospectral positive operators $B^+_{BS}(E)$ and $R^+(E)$ acting on $L^2( \mathbb R^+)$, with integral kernels given respectively by \eqref{2.8} and \eqref{2.23}, belong to the Hilbert-Schmidt class and  their Hilbert-Schmidt norms are given by:
\begin{equation}\label{2.24}
||B^+_{BS}(E)||_{T^2} = ||R^+(E)||_{T^2} = \sqrt{\frac{1}{6|E|}} \, \lambda \, \frac{\pi}{2}\,.
\end{equation}
Similarly, for any $\lambda>0,\, E< 0$, the isospectral positive operators $B_{BS}(E)$ and $R(E)$ acting on $L^2(\mathbb R)$, with integral kernels given respectively by \eqref{2.9} and \eqref{2.22}, belong to the Hilbert-Schmidt class and  their Hilbert-Schmidt norms are given by:
\begin{equation}\label{2.25}
||B_{BS}(E)||_{T^2}  = ||R(E)||_{T^2} = \sqrt{\frac{1}{6|E|}}  \, \lambda\, \pi \,.
\end{equation}
\end{theor}
In the Appendix, we shall introduce two new operators with similar mathematical properties to those studied in the present section.

\subsection{On the number of bound states}

At this stage we wish to determine an upper limit for the number of bound states, $N_E(H)$, for the Hamiltonian $H_{D^+}:= \left( -\frac{d^2}{dx^2} \right)_{D^+} - \frac \lambda x$ within the interval $(-\infty,E]$ with $E<0$. This estimation is based on the proof of Theorem XIII.10 in \cite{RSIV}, which gives an upper bound (the   B-S   bound) for the number of bound states for a Hamiltonian of the form $H= H_0 -V= -\nabla^2-V(\mathbf x)$, where $V(\mathbf x) \geq 0$ belongs to the Rollnik class \cite{RSIV}. The proof provided in \cite{RSIV} is based on the fact that, being $V(\mathbf x)$ a Rollnik potential, its   B-S   operator is Hilbert-Schmidt and this proves the   B-S   bound (and in fact, the Rollnik property is just a condition needed to get the Hilbert-Schmidt property of the relevant B-S operator, which is crucial for the existence of the bound in that context, and is irrelevant for our case). The result in \cite{RSIV} shows that $N_E(H)\le ||B||^2_{T^2}$.

The proof of this result can be mimicked for the case under our study, obtaining the same result. Thus, after \eqref{2.23}, we have the following bound  
\begin{equation}\label{2.25}
N_E(H_{D^+}) \le \frac{\lambda^2\, \pi^2}{24|E|}\,.
\end{equation}
In particular for $E=-1/n^2$, we have 
\begin{equation}\label{2.26}
N_{-1/n^2}(H_{D^+}) \le \frac{\lambda^2\,\pi^2 n^2}{24}\,.
\end{equation}
If instead of $H_{D^+}$, we have $H_D := \left( -\frac{d^2}{dx^2} \right)_{D} - \frac \lambda{|x|}$, so that, by using \eqref{2.24}, we obtain the following bounds:
\begin{equation}\label{2.27}
N_E(H_D) \le \frac{\lambda^2\, \pi^2}{6|E|}\,, \qquad   N_{-1/n^2}(H_D) \le \frac{\lambda^2\, \pi^2 n^2}{6}\,.
\end{equation}

\section{On the approximation of the one-dimensional Coulomb potential by a smeared potential}

Throughout the present section, we intend to revisit an issue that has been rather controversial, that is to say the connection between the one-dimensional Hamiltonian
\begin{equation}\label{3.1}
H_D(\lambda) := \left( - \frac{d^2}{dx^2}  \right)_D - \frac{\lambda}{|x|}\,,
\end{equation}
and a family of Hamiltonians of the type
\begin{equation}\label{3.2}
H(\varepsilon,\lambda) := \left( - \frac{d^2}{dx^2}  \right)_D - \lambda\, V_\varepsilon(|x|)\,, \qquad 0\le V_\varepsilon(|x|) \le  \frac 1{|x|}\,.
\end{equation}
Here, $V_\varepsilon(|x|)$ denotes a family of potentials in $L^2(\mathbb R)$, dependent on the positive parameter $\varepsilon>0$,  such that in the pointwise (a.e.) sense (see \cite{Fritz, KLA}) 
\begin{equation}\label{3.3}
\lim_{\varepsilon \to 0^+}V_\varepsilon(|x|) = \frac 1{|x|}\,.
\end{equation}
There are several choices in the literature for the family of potentials $V_\varepsilon(|x|)$. Among the most usual ones, we may cite $\frac 1{|x|+\varepsilon}$ and $\frac 1{\sqrt{x^2+\varepsilon^2}}$. 

In the sequel, we consider $E<0$ and $\varepsilon>0$. Let $B_\varepsilon(E)$ and $R_\varepsilon(E)$ two families of integral operators depending on the parameter $\varepsilon$. We define these operators by means of their kernels, which are, respectively,
\begin{equation}\label{3.4}
B_\varepsilon (x,y;|E|) =  \sqrt{V_\varepsilon(|x|)} \, \left[ \left( - \frac{d^2}{dx^2}  \right)_D + |E|  \right]^{-1}\!\!\! (x,y) \  \sqrt{V_\varepsilon(|y|)}\,,
\end{equation}
for $B_\varepsilon$, and for $R_\varepsilon$
\begin{equation}\label{3.5}
R_\varepsilon (x,y;|E|) =  \int_{-\infty}^{\infty}\left[\left(-\frac {d^2}{dx^2}\right)_{D}\!\!+ |E|\right]^{-\frac 12}\!\!\!(x,z)\,  V_{\epsilon}(|z|)\, \left[\left(-\frac {d^2}{dx^2}\right)_{D}\!\!+ |E|\right]^{-\frac12}\!\!\!(z,y)\, dz \,.
\end{equation}

\begin{theor}
We have the following operator limits on the Hilbert-Schmidt norm:
\begin{equation}\label{3.6}
\lim_{\varepsilon \to 0^+}  B_\varepsilon = B_{BS} \,, \qquad  \lim_{\varepsilon \to 0^+}  R_\varepsilon =R\,,
\end{equation}
where $B_{BS}$ and $R$ have been defined in \eqref{5} and in \eqref{2.20}, respectively. 
\end{theor}

\noindent
{\bf Proof}.- We first show that $B_\varepsilon$ is Hilbert-Schmidt for any $\varepsilon>0$. To accomplish this goal, we need only prove that its kernel is square integrable. In fact,
\begin{eqnarray}\label{3.7}
&&\hskip-0.8cm \int_{\mathbb R^2}   [B_\varepsilon (x,y;|E|)]^2\, dx\, dy   = \frac 1{4|E|} \int_{\mathbb R^2}  \, V_{\epsilon}(|x|)\, \left[e^{- |E|^{1/2} |x-y|} - e^{- |E|^{1/2}(|x|+|y|)}\right]^2\,  V_{\epsilon}(|y|)\, dx \, dy \nonumber \\ [2ex] 
  & &      \leq    \frac 1{4|E|}  \int_{\mathbb R^2}  V_{\epsilon}(|x|)\, \left[e^{- 2|E|^{1/2} |x-y|}+e^{- 2|E|^{1/2}(|x|+|y|)}\right]\,  V_{\epsilon}(|y|)\, dx \, dy \nonumber \\ [2ex] 
  &&     = \frac 1{4|E|}  \int_{\mathbb R^2}  V_{\epsilon}(|x|)\, e^{- 2|E|^{1/2} |x-y|}\, V_{\epsilon}(|y|)\, dx \, dy + \frac 1{4|E|} \left[\int_{-\infty}^{\infty} e^{- 2|E|^{1/2} |x|}\, V_{\epsilon}(|x|) \, dx \right]^2 \,. 
\end{eqnarray}

Now, let us use the notation $||-||_p$ for the norm in $L^p(\mathbb R)$, with $p=1,2$ and the star $\ast$ to denote convolution. Then the sum in the last row in \eqref{3.7} is nothing else but
\begin{eqnarray}\label{3.8}
\!\!&\! \!&\hskip-3cm \frac 1{4|E|}\, \left \| V_{\epsilon}\, \cdot \left(e^{- 2|E|^{1/2} |\cdot|}\, * V_{\epsilon} \right) \right \| _1 + \frac 1{4|E|}\, \left \| e^{- 2|E|^{1/2} |\cdot|}\, V_{\epsilon}  \right \| _1^2  \nonumber \\ [2ex] 
\!\!&\!\le \!&\!\!
 \frac 1{4|E|}\, \left \| e^{- 2|E|^{1/2} |\cdot|}\, \right \| _1\, \left \| V_{\epsilon}\, \right \| _2^2\, + \frac 1{4|E|}\, \left \| e^{- 2|E|^{1/2} |\cdot|}\, \right \| _2^2\, \left \| V_{\epsilon}\, \right \| _2^2 \nonumber \\ [2ex] 
\!\!&\!=\!&\!\! 
\frac {\left \| V_{\epsilon}\, \right \| _2^2}{4|E|}\, \left(\frac 1{|E|^{1/2}}\, + \frac 1{2|E|^{1/2}}  \right)=\, \frac {3\, \left \| V_{\epsilon}\, \right \| _2^2}{8|E|^{3/2}}\,,
\end{eqnarray}
where the inequality in \eqref{3.8} is a consequence of Young's inequality \cite{RSII} for the convolution in addition to the well-known one due to Schwartz. This shows that $B_\varepsilon$ is Hilbert-Schmidt for all $\varepsilon>0$. A similar proof can be produced in order to establish the analogous property for the family $R_\varepsilon$. 

Next, we prove that the limit \eqref{3.6} holds in the weak operator sense. Let $\psi(x),\varphi(x)\in \mathcal S_0$, where $\mathcal S_0$ is the space of all Schwartz functions that vanish at the origin. The space $\mathcal S_0$ is dense in $L^2(\mathbb R)$. Let us consider the following expression, where we omit the factor $1/(2|E|^{1/2})$ and denote by $[-]$ the central factor in \eqref{2.6} or \eqref{3.7} and $V(|x|) := 1/|x|$:
\begin{eqnarray}\label{3.9}
\langle \psi|(B_\varepsilon - B_{BS})\varphi\rangle 
\!\!&\! = \!&\!\!
 \int_{\mathbb R^2}  V_\varepsilon^{1/2}(|x|) \psi^*(x) [-] V_\varepsilon^{1/2}(|y|) \varphi(y)\, dx\, dy  \nonumber \\  [1ex]  
  \!\!&\!  \!&\!\! 
-  \int_{\mathbb R^2}  V^{1/2}(|x|) \psi^*(x) [-] V^{1/2}(|y|) \varphi(y)\, dx\, dy  
 \nonumber \\  [2ex] 
  \!\!&\!   \!&\!\! 
+  \int_{\mathbb R^2}  V_\varepsilon^{1/2}(|x|) \psi^*(x) [-] V^{1/2}(|y|) \varphi(y)\, dx\, dy \nonumber \\ [2ex] 
 \!\!&\!   \!&\!\! 
 -  \int_{\mathbb R^2}  V_\varepsilon^{1/2}(|x|) \psi^*(x) [-] V^{1/2}(|y|) \varphi(y)\, dx\, dy  \nonumber \\ [2ex] 
  \!\!&\!  = \!&\!\! 
 \int_{\mathbb R^2}  V_\varepsilon^{1/2}(|x|)  \psi^*(x) [-] [V_\varepsilon^{1/2}(|y|) - V^{1/2}(|y|)] \varphi(y)\, dx\, dy  \nonumber \\  [2ex] 
 \!\!&\!   \!&\!\! 
-  \int_{\mathbb R^2}  [V_\varepsilon^{1/2}(|x|) - V^{1/2}(|x|) ] \psi^*(x) [-] V^{1/2}(|y|)] \varphi(y)\, dx\, dy \,. \qquad
\end{eqnarray}
Since $[-] \le 2$, the modulus of the fifth row in \eqref{3.9} is bounded by (omitting the 2)
\begin{eqnarray}\label{3.10}
\left[  \int_{-\infty}^\infty V_\varepsilon^{1/2}(|x|)  |\psi(x)|\, dx \right]   \left[  \int_{-\infty}^\infty  |V_\varepsilon^{1/2}(|y|) - V^{1/2}(|y|)|\, | \varphi(y)|\, dy  \right]\,.
\end{eqnarray}
As $V^{1/2}_\varepsilon (|x|) \le V^{1/2}(|x|)$ for all $\varepsilon>0$, we may apply the dominated convergence theorem to both factors in \eqref{3.10} so that this product vanishes as $\varepsilon \to 0$. The same argument goes for the last term in \eqref{3.9}. Now, both operators $B_\varepsilon$ and $B_{BS}(E)$ are bounded and $\mathcal S_0$ is dense, which proves that $B_\varepsilon \to B_{BS}(E)$ in the weak operator sense. 

Given that $B_\epsilon(x,y;|E|) \le K_{BS}(x,y;|E|)$ and $\lim_{\varepsilon \to 0} B_\epsilon(x,y;|E|) = K_{BS}(x,y;|E|)$ pointwise, we may also apply the dominated convergence theorem in order to obtain:
\begin{equation}\label{3.11}
\lim_{\varepsilon \to 0^+} ||B_\varepsilon||^2_{T_2} = \lim_{\varepsilon \to 0^+}    \int_{\mathbb R^2} ^\infty [B_\epsilon(x,y;|E|)]^2\, dx\, dy = \int_{-\infty}^\infty [K_{BS}(x,y;|E|)]^2\, dx\, dy = ||B_{BS}||^2_{T_2}\,,
\end{equation}
so that $\lim_{\varepsilon \to 0^+} ||B_\varepsilon||_{T_2} = ||B_{BS}||_{T_2}$. Then, by invoking Theorem 2.21 in \cite{S Trace}, we conclude that
\begin{equation}\label{3.12}
\lim_{\varepsilon \to 0^+} ||B_\varepsilon - B_{BS}||_{T_2} =0\,, 
\end{equation}
which proves the first assertion in \eqref{3.6}. The second one is a consequence of the isospectrality of $B_\varepsilon$ and $R_\varepsilon$ on one side and $B_{BS}$ and $R$ on the other. \hfill $\blacksquare$

\begin{coro}
The resolvent of the operator
\begin{equation}\label{3.13}
H_D(\varepsilon,\lambda) := \left( -\frac{d^2}{dx^2} \right)_D -\lambda\, V_\varepsilon(|x|)\,,
\end{equation}
converges in the norm topology of bounded operators to the resolvent of the operator
\begin{equation}\label{3.14}
H_D(\lambda) := \left( -\frac{d^2}{dx^2} \right)_D - \frac{\lambda}{|x|}\,.
\end{equation}
\end{coro}

\medskip
\noindent
{\bf Proof}.- The fact that $\lim_{\varepsilon \to 0^+} R_\varepsilon = R$ in the Hilbert-Schmidt norm implies that
\begin{eqnarray}\label{3.15}
\lim_{\varepsilon \to 0^+} (H_D- E)^{-1/2} \left(H_D(\varepsilon,\lambda)  - H_D(\lambda) \right) (H_D -E)^{-1/2}  =\qquad \nonumber \\ [2ex] 
= \lim_{\varepsilon \to 0^+} (H_D- E)^{-1/2} \left( \frac{\lambda}{|x|}\, -\lambda\, V_\varepsilon(|x|) \right) (H_D -E)^{-1/2} \,,
\end{eqnarray}
in the Hilbert-Schmidt norm. Since for any H-S operator $A$, $||A|| \le ||A||_{T^2}$, where $||-||$ is the norm for bounded operators, then, the above limit holds for this norm as well. Then, we use the proof of part (c) of Theorem VIII.25 in \cite{RSI} to conclude that $H_D(\varepsilon,\lambda) \to H_D(\lambda)$ in the norm resolvent sense\footnote{This result provides a sufficient condition for a sequence of positive self-adjoint operators with a common form domain, converging in norm to a given operator $A$, to converge in the norm resolvent sense to $A$.}. \hfill $\blacksquare$

\medskip

 As a consequence of the above corollary, the remarkable functional analytic result due to Klaus in \cite{KLA} can be achieved for the Hamiltonians with our approximating potentials without introducing the upper bound $W(x)$, appearing in his requirement (i) and needed to establish (13) in his paper, since in our case the compactness of the limiting B-S operator $B_{BS}$ is ensured by Theorem 1 in Section 2. Then, Klaus' proof of the convergence of the operator $H(\epsilon,\lambda)=-\frac {d^2}{dx^2}- \lambda V_{\epsilon}(|x|)$ (note that in this Hamiltonian we do not use the Dirichlet boundary conditions) to $\left(-\frac {d^2}{dx^2}\right)_{D}-\frac {\lambda}{|x|}$  in the norm resolvent sense, based on a suitable expansion of the resolvent of $H(\epsilon,\lambda)$ (see (15)-(17) in \cite{KLA}), can be mimicked word by word taking account of the following:

i) the difference between the free resolvent and the one involving the Dirichlet Laplacian is given by a rank one operator, namely
\begin{equation}\label{3.16}
\left(-\frac {d^2}{dx^2}+|E|\right)^{-1}-\left[\left(-\frac {d^2}{dx^2}\right)_{D}+|E|\right]^{-1}=\, \frac 1{2|E|^{1/2}}\,|e^{-|E|^{1/2}\,|\cdot | }\,\rangle\, \langle e^{-|E|^{1/2}\,|\cdot | }\,|=P_{|E|}\,,
\end{equation}

ii) the resolvent of the positive rank one operator $P_{|E|}$ is given by
\begin{eqnarray}\label{3.17}
\left(P_{|E|}+|E| \right)^{-1} \!&\!=\!& \! \frac1{|E|} - \frac 1{|E|\left[\left \|e^{-|E|^{1/2}\,|\cdot |}\, \right \|_2^2+|E|\right]}\,P_{|E|}  \nonumber \\ [1.2ex] 
 \!&\!=\!& \! \frac 1{|E|} - \frac 1{|E|^{1/2}\left[ 1+|E|^{3/2}\right]}\,P_{|E|},
\end{eqnarray}

iii) apart from assumption (4) in \cite{KLA}, replaced by the upper bound in \eqref{3.2}, our approximating potentials $-\lambda\, V_{\epsilon}(|x|)\,$ clearly satisfy all the other requisites (5)-(8) in that paper, so that the crucial step following (18) holds in our case as well.

Hence, our result can be stated as a formal theorem in the following form:

\begin{theor}
Let  $0\leq V_{\epsilon}(|x|)\,\leq \frac 1{|x|}$ be a square summable potential such that $V_{\epsilon}(|x|)\, \rightarrow \frac 1{|x|}$ as $\epsilon \rightarrow 0^+$ in the pointwise (a.e.) sense. Then, for any $\lambda >0$, the self-adjoint Hamiltonian $H(\epsilon,\lambda)=-\frac {d^2}{dx^2}- \lambda V_{\epsilon}(|x|)$ converges in the norm resolvent sense to the self-adjoint Hamiltonian $H_{D}(\lambda)=\left[-\frac {d^2}{dx^2}\right]_{D}-\frac {\lambda}{|x|}$. 
\end{theor}

\section{Concluding Remarks}

The one-dimensional Schr\"odinger equation with Coulomb potential has been the object of various studies. Differently from its three-dimensional version,  no unique self-adjoint realisation of the Hamiltonian pertaining to this equation exists. In the present study, we have chosen one of such realisations which fixes the Hamiltonian, $H_D(\lambda)$, characterised by the Dirichlet boundary conditions at the origin. The parameter $\lambda$ is just a multiplicative constant that appears in the potential. Once we have made this choice, there are still some open problems concerning this potential. We have discussed two of them.

In the first place, we have obtained the   B-S   operator for the given Hamiltonian. The   B-S   operator is intrinsically related with the Hamiltonian and, under some conditions, provides a tool to obtain bound states and scattering resonances. A sufficient condition thereof is that the   B-S   operator be Hilbert-Schmidt. We have rigorously shown that, in fact, this operator is Hilbert-Schmidt.

It is interesting to remark that in the process of our proof, we have obtained an integral expression for $\zeta(2)$, where $\zeta(s)$ represents the Riemann zeta function, which, at least to the best of our knowledge, seems to be new. This is
\begin{equation}\label{51}
\int_0^\infty\int_0^\infty e^{-x} \left[ \frac{e^{\min(x,y)}-1}{x^{1/2}\, y^{1/2}} \right]  e^{-y}\, dx\, dy = \sum_{\ell=1}^\infty \frac{1}{\ell^2} = \zeta(2) = \frac{\pi^2}{6}\,.
\end{equation}

The second issue analysed here deals with possible regularisations of the Hamiltonian based on approximations of the one-dimensional Coulomb potential by another potential which depends on a certain parameter. As this parameter goes to zero, the approximating potential approaches the Coulomb one. We have shown that such approximations make sense since the   B-S   operator of the approximating Hamiltonian converges to the   B-S   operator of the exact Hamiltonian in the Hilbert-Schmidt norm. Furthermore, this implies the convergence of these approximating Hamiltonians to  the one with the negative Dirichlet Laplacian perturbed by the attractive one-dimensional Coulomb potential in the norm resolvent sense.

A thorough analysis of the discrete spectrum of $H_D(\lambda)$ will be carried out in a forthcoming paper.

\appendix

\section{Appendix}

Intimately linked and with a similar shape to the   B-S  operators $B_{BS}(E)$ and $B^+_{BS}(E)$ are the integral operators $b(E)$ and $b^+(E)$ also defined on $L^2(\mathbb R)$ and $L^2(\mathbb R^+)$, respectively. In principle, these operators have a purely mathematical interest: the traces of the operators given by the kernels $B_{BS}(E)$ and $b(E)$ coincide on given subspaces of $L^2(\mathbb R)$ and the same happens with $B^+_{BS}(E)$ and $b^+(E)$ on $L^2(\mathbb R^+)$. These operators are defined through their respective kernels as follows:
\begin{equation}\label{5.1}
b(x,y;|E|):= \frac{e^{-|E|^{1/2}\, |x|}}{|x|^{1/2}} \frac{e^{2|E|^{1/2}\, |x|^{1/2}\, |y|^{1/2}} - 1 }{2|E|^{1/2}}  \frac{e^{-|E|^{1/2}\, |y|}}{|y|^{1/2}} \,,
\end{equation}
and
\begin{equation}\label{5.2}
b^+(x,y;|E|):= \frac{e^{-|E|^{1/2}\, x}}{x^{1/2}} \frac{e^{2|E|^{1/2}\, x^{1/2}\, y^{1/2}} - 1 }{2|E|^{1/2}}  \frac{e^{-|E|^{1/2}\, y}}{y^{1/2}}\,, 
\end{equation}
where $x\ge 0$ and $y\ge0$ in \eqref{5.2}. 
It is noteworthy to compare  \eqref{5.1} with  \eqref{2.7} and \eqref{5.2} with \eqref{2.6}.

The comparison between these operators is simple. In fact, taking account of $\min(x,y)\, \le  x^{1/2}\, y^{1/2}$ for any $x,y\ge 0$, one has that for any $\psi(x) \in L^2(\mathbb R^+)$:
\begin{equation}\label{5.3}
0 \le \left(\psi, B^+_{BS}\,\psi\, \right) \le \left(\psi, b^+\,\psi\, \right),
\end{equation}
where $\left( -,-\right)$ denotes the scalar product between two vectors in the given Hilbert space. This means that $0 \le B^+_{BS} \le b^+$. Similarly, $0 \le B_{BS} \le b$. 

The reason behind our interest in these new operators is that their trace coincides with that of their counterparts, when we consider the restriction of all these operators on given subspaces of $L^2(\mathbb R^+)$, for $b^+$ and $B^+_{BS}$, or $L^2(\mathbb R)$ for $b$ and $B_{BS}$.
We have already commented on the reason to consider these two new operators.  
In fact, take $L^2[0,N]$ considered as a subspace of $L^2(\mathbb R^+)$. On the interval $[0,N]$, we have
\begin{equation}\label{5.4}
\int_0^N B^+(x,x;|E|)\, dx = \int_0^N b^+(x,x;|E|)\, dx = \frac{1}{2|E|^{1/2}}  \int_0^N \frac{1- e^{-2|E|^{1/2}\, x}}{x}\, dx\,.
\end{equation}
The last integral is obviously convergent, as the integrand has a removable singularity at the origin. Thus, the traces of both restrictions coincide and diverge in the limit $N \to \infty$. This is a manifestation of the well-known infrared divergence, i.e., divergence for large distances or, equivalently, short momenta, of the one-dimensional Coulomb potential. The same can be said about the restrictions of $b$ and $B_{BS}$ to the spaces $L^2[-N,N]$. 

It is interesting to show that, although they are not trace class,  both $b^+$ and $b$ are Hilbert-Schmidt operators on $L^2(\mathbb R^+)$ and $L^2(\mathbb R)$. Both are, rather obviously, positive operators.  We just give the proof for $b^+$, the proof for $b$ being similar. We need only compute the following integral:
\begin{eqnarray}\label{5.5}
\int_0^\infty \int_0^\infty [b^+(x,y;|E|)]^2\, dx dy   \!\!&\!  = \!&\!\! 
\frac{1}{4|E|}  \int_0^\infty \int_0^\infty e^{-2|E|^{1/2}\, x} \left[ \frac{e^{2|E|^{1/2}\, x^{1/2}\, y^{1/2}} -1}{x^{1/2}\, y^{1/2}}  \right]^2  e^{-2|E|^{1/2}\, y}\, dx dy \nonumber \\ [2ex] 
  \!\!&\!  = \!&\!\! 
 \frac{1}{4|E|}  \int_0^\infty \int_0^\infty e^{-x} \left[ \frac{e^{x^{1/2}\, y^{1/2}} -1 }{x^{1/2}\, y^{1/2}} \right]^2 e^{-y}\,dx  dy\,.
\end{eqnarray}

Now, let us use the well-known expansion of $\frac{e^x-1}{x}$ in the last integral in \eqref{5.5}. Then, this integral is equal to:
\begin{eqnarray}\label{5.6}
\frac{1}{4|E|}  \int_0^\infty \int_0^\infty e^{-x} \left[\sum_{\ell=0}^\infty \frac{x^{\ell/2}\, y^{\ell/2}}{(\ell+1)!}  \right]^2 e^{-y}\,dx  dy   
  \!\!&\!  = \!&\!\! 
\frac{1}{4|E|} \sum_{\ell=0}^\infty \sum_{m=0}^\infty \frac{1}{(\ell+1)! (m+1)!} \left[ \int_0^\infty x^{\frac{\ell+m}{2}} e^{-x} dx \right]^2  \nonumber \\ [2ex] 
  \!\!&\!  = \!&\!\! 
 \frac{1}{4|E|} \sum_{\ell=0}^\infty \sum_{m=0}^\infty \frac{\Gamma^2 \left( \frac{\ell+m}{2} +1 \right)}{(\ell+1)! (m+1)!} = \frac{13 \pi^2}{72|E|}\,.
\end{eqnarray}
Expressions \eqref{5.5} and \eqref{5.6} give the Hilbert-Schmidt norm of $b^+$, which is
\begin{equation}\label{5.7}
||b^+||_{T^2} = \sqrt{\frac{13}{2|E|}} \, \frac\pi 6\,.
\end{equation}
Similar manipulations give the Hilbert-Schmidt norm of $b$, which is given by
\begin{equation}\label{5.8}
||b||_{T^2} = \sqrt{\frac{13}{2|E|}} \, \frac\pi 3 \,.
\end{equation}
All operators considered here are positive and Hilbert-Schmidt, so that the squares of their Hilbert-Schmidt  trace norms is the sum of the squares of their respective eigenvalues. Let us denote  the eigenvalues of $b^+$ and $B^+_{BS}$ by $\{\lambda_n^+(|E|)\}$ and by $\{\Lambda^+_n(|E|)\}$, respectively. We obtain the following results:
\begin{equation}\label{5.9}
\sum_{n=1}^\infty \Lambda_n^+(|E|) = \sum_{n=1}^\infty \lambda^+_n(|E|) = \infty\,,
\end{equation}
and 
\begin{equation}\label{5.10}
\sum_{n=1}^\infty [\Lambda_n^+(|E|)]^2 = \frac{\pi^2}{24|E|} = \frac{3}{13} \sum_{n=1}^\infty [\lambda^+_n(|E|)]^2\,.
\end{equation}

Then, let us denote the eigenvalues of $b$ and $B_{BS}$ by $\{\lambda_n(|E|)\}$ and by $\{\Lambda_n(|E|)\}$, respectively. Now, we have that
\begin{equation}\label{5.11}
\sum_{n=1}^\infty \Lambda_n(|E|) = \sum_{n=1}^\infty \lambda_n(|E|) = \infty\,,
\end{equation}
and 
\begin{equation}\label{5.12}
\sum_{n=1}^\infty [\Lambda_n(|E|)]^2 = \frac{\pi^2}{6|E|} = \frac{3}{13} \sum_{n=1}^\infty [\lambda_n(|E|)]^2\,.
\end{equation}
Observe that the relations are identical in both cases. On the basis of the considerations in this Appendix, we wonder whether it might be worth investigating  further such pairs of positive compact operators along the lines of previous research on diagonals of positive trace class operators (see \cite{Arve}).

\section*{Acknowledgements}

Firt of all, we wish to thank the anonymous referees whose constructive criticism has led to significant improvements of our manuscript.
This research was supported by the Q-CAYLE project, funded by the European Union-Next Generation UE/MICIU/Plan de Recuperacion, Transformacion y Resiliencia/Junta de Castilla y Leon (PRTRC17.11), and also by RED2022-134301-T and PID2023-148409NB-I00, financed by MI-CIU/AEI/10.13039/501100011033.  
Financial support of the Department of Education of the Junta de Castilla y León and FEDER Funds is also gratefully acknowledged (Reference: CLU-2023-1-05).
S. Fassari wishes to express his heartfelt thanks to Prof. L.M. Nieto and to Prof. M. Gadella for making his stay at their institution, Department of Theoretical, Atomic and Optical Physics of the University of Valladolid, during the second half of April 2023 possible through the aforementioned funding sources, as well as to all the other members of the department for their warm hospitality. Also, M. Gadella wishes to express his gratitude to Prof. J. T. Lunardi and his group for their warm hospitality at the University of Ponta Grossa.


\begin{thebibliography}{99}
\bibitem{F96} S. Fassari, A note on the eigenvalues of the Hamiltonian of the harmonic oscillator perturbed by the potential $\lambda x^2/(1+gx^2)$,  {\it Rep. Math. Phys.} {\bf 37} (1996), 283-293.
\end{thebibliography}
\end{document}